\pdfoutput=1
\documentclass[12pt, letterpaper]{article}

\usepackage{amsmath}
\usepackage{amsfonts}
\usepackage{amssymb}
\usepackage{graphicx}
\usepackage{color}

\begin{document}

\begin{flushright} {\footnotesize YITP-20-72, IPMU20-0058}  \end{flushright}
\vspace{5mm}
\vspace{0.5cm}
\begin{center}

\def\thefootnote{\fnsymbol{footnote}}

{\Large {\bf Cosmology and gravitational waves in consistent $D\to 4$ Einstein-Gauss-Bonnet gravity}}
\\[1cm]

{Katsuki Aoki$^{1}\footnote{katsuki.aoki@yukawa.kyoto-u.ac.jp}$, Mohammad Ali Gorji$^{1}\footnote{gorji@yukawa.kyoto-u.ac.jp}$, 
 Shinji Mukohyama$^{1,2}\footnote{shinji.mukohyama@yukawa.kyoto-u.ac.jp}$}
\\[.7cm]

{\small \textit{$^1$Center for Gravitational Physics, Yukawa Institute for Theoretical
		Physics, Kyoto University, 606-8502, Kyoto, Japan
}}\\

{\small \textit{$^2$Kavli Institute for the Physics and Mathematics of the Universe (WPI), The University of Tokyo, 277-8583, Chiba, Japan}}\\

\end{center}

\vspace{.8cm}

\hrule \vspace{0.3cm}

\begin{abstract} 
In a very recent paper \cite{Aoki:2020lig}, we have proposed a novel $4$-dimensional gravitational theory with two dynamical degrees of freedom, which serves as a consistent realization of $D\to4$ Einstein-Gauss-Bonnet gravity with the rescaled Gauss-Bonnet coupling constant $\tilde{\alpha}$. This has been made possible by breaking a part of diffeomorphism invariance, and thus is consistent with the Lovelock theorem. In the present paper, we study cosmological implications of the theory in the presence of a perfect fluid and clarify the similarities and differences between the results obtained from the consistent $4$-dimensional theory and those from the previously considered, naive (and inconsistent) $D\rightarrow 4$ limit. Studying the linear perturbations, we explicitly show that the theory only has tensorial gravitational degrees of freedom (besides the matter degree) and that for $\tilde{\alpha}>0$ and $\dot{H}<0$, perturbations are free of any pathologies so that we can implement the setup to construct early and/or late time cosmological models. Interestingly, a $k^4$ term appears in the dispersion relation of tensor modes which plays significant roles at small scales and makes the theory different than not only general relativity but also many other modified gravity theories as well as the naive (and inconsistent) $D\to 4$ limit. Taking into account the $k^4$ term, the observational constraint on the propagation of gravitational waves yields the bound $\tilde{\alpha} \lesssim (10\,{\rm meV})^{-2}$. This is the first bound on the only parameter (besides the Newton's constant and the choice of a constraint that stems from a temporal gauge fixing) in the consistent theory of $D\to 4$ Einstein-Gauss-Bonnet gravity. 
\end{abstract}
\vspace{0.5cm} 
\hrule
\def\thefootnote{\arabic{footnote}}
\setcounter{footnote}{0}

\newpage

\section{Introduction}\label{introduction}

According to the Lovelock theorem \cite{Lovelock:1971yv,Lovelock:1972vz}, any rank-2 divergence-free tensor constructed from the metric and its first two derivatives in $4$ spacetime dimensions is a linear combination of the Einstein tensor and the metric itself. Therefore, in 4 spacetime dimensions any covariant action made of the metric and its derivatives that leads to second order equations of motion is the Einstein-Hilbert action with or without a cosmological constant up to a boundary term. This is a mathematically established theorem and serves as a part of the common knowledge in the research community of gravitational theories. For example, one can in principle add the Gauss-Bonnet term to the $4$-dimensional action but it is a boundary term and thus does not contribute to the equations of motion.

Nonetheless, it was recently claimed that if we start with Einstein-Gauss-Bonnet gravity in $D$ spacetime dimensions with $D>4$ and if take the $D\to 4$ limit with $\tilde{\alpha} = (D-4)\alpha$ kept finite, where $\alpha$ is the Gauss-Bonnet coupling constant, non-trivial contributions from the Gauss-Bonnet term should remain in the form of covariant corrections to the Einstein equation \cite{Glavan:2019inb}. This claim explicitly contradicts with the Lovelock theorem\footnote{The paper \cite{Glavan:2019inb} claims that the $D\to 4$ limit should be taken at the level of equations of motion, rather than at the level of the action (or Hamiltonian). Note, however, that the Lovelock theorem is a statement about the equations of motion. Therefore, independently from the existence/non-existence of the action, the claim of \cite{Glavan:2019inb} is anyway in explicit contradiction with the Lovelock theorem. See discussions in \cite{Aoki:2020lig} for the $D\to 4$ limit at the level of both the equations of motion and the Hamiltonian.}.

Some subtleties of this limit were then revealed by many papers. For instance, by taking the $D\to4$ limit directly from a $D$-dimensional spacetime, it is shown that an extra scalar degree of freedom (dof), originating from the extra $(D-4)$-dimensional space, shows up \cite{Lu:2020iav,Kobayashi:2020wqy,Bonifacio:2020vbk}. This result was also confirmed from another perspective \cite{Fernandes:2020nbq,Hennigar:2020lsl} through adding counter terms in $D$ dimensions \cite{Mann:1992ar}. At the level of equations of motion, it was shown that, if we keep all conditions of the Lovelock theorem but the divergence-freedom (i.e. the Bianchi identity), the $D\to 4$ limit inevitably ends up with breaking the Bianchi identity \cite{Gurses:2020ofy} and therefore it is not diffeomorphism invariant. See also other papers \cite{Ai:2020peo,Mahapatra:2020rds,Arrechea:2020evj} in this direction.

Moreover, the $D\to 4$ limit is not unique~\cite{Aoki:2020lig} so that one should choose the way of taking the limit to consistently describe 4-dimensional spacetimes. Indeed, depending on the choice of the ``regularisation scheme'', many theories with different number of degrees of freedom and different properties would arise. In particular and in the simplest case, the scalar-tensor description of the $D\to 4$ limit of the Einstein-Gauss-Bonnet gravity proposed by \cite{Lu:2020iav,Kobayashi:2020wqy,Fernandes:2020nbq,Hennigar:2020lsl} admits a 4-dimensional theory which belongs to the Horndeski theory. However, the scalar field lacks the quadratic kinetic term that signals the infinite strong coupling problem. When one uses the canonically normalized scalar field before taking the limit, the kinetic term can be retained after taking the limit but divergent non-linear interactions arise which again signals that the system is strongly coupled \cite{Bonifacio:2020vbk}. Around the infinitely strongly coupled background, the cutoff of the effective field theory is zero, meaning that a UV completion is required to make any predictions. One cannot describe any physical processes without knowledge of the UV completion, namely quantum gravity. Since this scalar-tensor theory is just a classical gravitational theory, the theory is unable to make any predictions in a reliable way when it is infinitely strongly coupled. The existence of the infinite strong coupling is explicitly confirmed around FLRW backgrounds \cite{Kobayashi:2020wqy} (see also \cite{Bonifacio:2020vbk,Ma:2020ufk}). Another scalar-tensor description without the strong coupling was proposed by \cite{Bonifacio:2020vbk} under a different scaling limit of the Gauss-Bonnet coupling which is therefore different than what was originally proposed in \cite{Glavan:2019inb}. The resultant theory is just the well-known $(\partial \phi)^4$ theory and is not new. In summary, apart from the non-uniqueness problem, the scalar-tensor descriptions of the $D\to 4$ limit of the Einstein-Gauss-Bonnet gravity are either a pathological strongly coupled theory or an already well-known theory with an extra degree of freedom. Although the cosmological and black hole solutions found by \cite{Glavan:2019inb} can be ``reproduced'' by the scalar-tensor realization of $D\to4$ Einstein-Gauss-Bonnet gravity \cite{Lu:2020iav,Kobayashi:2020wqy}, as stated, the theory cannot be trusted due to the infinite strong coupling problem \cite{Kobayashi:2020wqy} (the black hole spacetime must be also infinitely strongly coupled at least in the asymptotic region). Hence, these solutions {\it cannot} be realized by the scalar-tensor description in a consistent manner. To use the exact ``solutions'' of \cite{Glavan:2019inb}, one should consistently take the $D\to 4$ limit as we have investigated in \cite{Aoki:2020lig}. In other words, although the cosmological and black hole solutions that are obtained in the original paper \cite{Glavan:2019inb} are solutions of both the scalar-tensor realization suggested in \cite{Lu:2020iav,Kobayashi:2020wqy} and the consistent setup in \cite{Aoki:2020lig}, it is important to note that we can only trust them in the setup of \cite{Aoki:2020lig} since they are strongly coupled in the scalar-tensor description \cite{Lu:2020iav,Kobayashi:2020wqy}. In this sense, the setup in \cite{Aoki:2020lig} serves as the consistent realization of the $D\to4$ Einstein-Gauss-Bonnet gravity.

More precisely, in \cite{Aoki:2020lig}, we proposed a $4$-dimensional theory that serves as a consistent realization of the $D\rightarrow 4$ limit of the Einstein-Gauss-Bonnet gravity with two dofs. We first concluded that the $D\to4$ limit should either break a part of diffeomorphism or leads to extra dofs, in agreement with the Lovelock theorem. Therefore, a consistent $D\rightarrow 4$ theory with only two dofs, if exists, should not possess a $4$-dimensional diffeomorphism invariant description in terms of the metric and its derivatives only, contrary to the claim in \cite{Glavan:2019inb}, but in accordance with the Lovelock theorem \cite{Lovelock:1971yv,Lovelock:1972vz}. Second, we have shown that it is possible to construct a consistent $4$-dimensional theory with only two dofs in the context of the minimally modified gravity theories \cite{DeFelice:2015hla,Lin:2017oow,Aoki:2018zcv,Mukohyama:2019unx,DeFelice:2020eju}. We proposed a consistent $4$-dimensional theory realizing the idea of the $D\rightarrow 4$ Einstein-Gauss-Bonnet gravity by requiring the following five conditions:
\begin{enumerate}
\item \label{con1}
The theory is invariant under the 3-dimensional spatial diffeomorphism
\begin{equation}\label{spatial-diff}
x^i \to \tilde{x}^i(t,x^k)\,.
\end{equation}
\item \label{con2}
The number of the local physical dofs in the gravitational sector is two.
\item \label{con3}
The theory reduces to GR when the ``rescaled Gauss-Bonnet coupling'' $\tilde{\alpha}$ vanishes. 
\item \label{con4}
Each term in the correction to GR is 4th-order in derivatives.
\item \label{con5}
If the Weyl tensor of the spatial metric and the Weyl part of $K_{ik}K_{jl}-K_{il}K_{jk}$, where $K_{ij}$ is the extrinsic curvature, vanish for a solution of $D$-dimensional Einstein-Gauss-Bonnet gravity, then the $D\rightarrow 4$ limit of the solution is a solution of the $4$-dimensional theory. 
\end{enumerate}
The conditions \ref{con1}-\ref{con3} state that the $4$-dimensional theory is a Lorentz violating\footnote{Lorentz violation is in the gravity sector and suppressed by $\tilde{\alpha}$. Thus, as far as the
matter (i.e. the standard model) is minimally coupled to gravity, Lorentz violation in the matter sector induced by graviton loops is suppressed not only by $\tilde{\alpha}$ but also by negative power of $M_{\rm Pl}^2$. In this sense Lorentz violation in the theory is under control.} gravity with two dofs which has a continuous GR limit. The condition \ref{con4} then restricts the form of correction terms to those with the ``Gauss-Bonnet'' structure which is 4th-order in derivatives. The condition \ref{con5} finally determines the direct relation between the $D\to 4$ limit of the $D$-dimensional theory and the $4$-dimensional theory. It is this condition that makes it natural to call this theory a theory of $D\to 4$ Einstein-Gauss-Bonnet gravity. Note that the Lorentz violation is inevitable due to the Lovelock theorem and, in the present case, is induced by the counter terms that regularise the Hamiltonian or/and the action of the $D$-dimensional theory in the course of the $D\to 4$ limit. As we will review in Section \ref{sec_theory}, our theory is defined with an additional constraint that stems from a temporal gauge condition in the $D$-dimensional theory. As a consequence, the $\tilde{\alpha} \rightarrow 0$ limit of the $4$-dimensional theory is GR with the temporal gauge degree of freedom fixed by the additional constraint, and then there is no discontinuity between $\tilde{\alpha} \rightarrow 0$ and $\tilde{\alpha} =0$. As argued in \cite{Aoki:2020lig}, the conditions \ref{con1}-\ref{con5} uniquely determine the $4$-dimensional theory up to a choice of the ``gauge-fixing'' constraint.

The present paper is devoted to a study of cosmological implications of the model and clarification of the similarities and differences between results obtained from the consistent $4$-dimensional theory~\cite{Aoki:2020lig} and those from the naive (and inconsistent) $D\rightarrow 4$ limit~\cite{Glavan:2019inb}. In Section \ref{sec_theory}, we briefly review the consistent theory of $D\rightarrow 4$ Einstein-Gauss-Bonnet gravity with two dofs and explain our setup. We then study the background dynamics in spatially flat FLRW universe in Section \ref{sec_background}. We explicitly show that the background equations indeed coincide with those obtained by~\cite{Glavan:2019inb} since the background flat FLRW universe is spatially conformally flat. In Sections \ref{sec_scalar}, \ref{sec_vector}, and \ref{sec_tensor}, we then study the scalar, vector, and tensor sectors, respectively, of the linear perturbations. In particular, we find that the dispersion relation of the gravitational waves is modified by a $k^4$ term which is a direct evidence that the result in the consistent theory deviates from that in the naive (and inconsistent) $D\rightarrow 4$ limit. Section \ref{summary} is devoted to a summary and discussions.

\section{A consistent $D\to 4$ Einstein-Gauss-Bonnet gravity}
\label{sec_theory}

As mentioned in Introduction, the consistent $4$-dimensional theory with two dofs does not allow for a $4$-dimensionally covariant description in terms of the metric and its derivatives. A natural setup to write down the theory is then the so-called Arnowitt-Deser-Misner (ADM) formalism based on the lapse function $N$, shift vector $N^i$, and spatial metric $\gamma_{ij}$. We can then write the $4$-dimensional metric, that matter fields are minimally coupled to, as
\begin{equation}\label{metric}
ds^2 = g_{\mu\nu} dx^{\mu} dx^{\nu} = - N^2 dt^2 + \gamma_{ij} ( dx^i + N^i dt ) ( dx^j + N^j dt ) \,.
\end{equation}

For a generic choice of the additional constraint stemming from the gauge-fixing condition, the consistent $D\rightarrow 4$ Einstein-Gauss-Bonnet theory is defined in the Hamiltonian formalism~\cite{Aoki:2020lig}. Moreover, for a simple gauge condition, which is compatible with both cosmological and spherically symmetric backgrounds, we can explicitly perform the Legendre transformation and find the corresponding Lagrangian. Therefore, we first introduce the Lagrangian formalism based on this particular gauge condition in the next subsection and, in the subsection after that, we discuss about the role of a general gauge condition in the Hamiltonian framework.

\subsection{Lagrangian formalism}

In this subsection we suppose the simple and convenient ``gauge condition'' of the form 
\begin{equation}\label{gauge}
{}^3\! \mathcal{G}=\sqrt{\gamma}D_k D^k(\pi^{ij}\gamma_{ij}/\sqrt{\gamma}) \approx 0 \,,
\end{equation}
where $\pi^{ij}$ is the canonical momentum conjugate to $\gamma_{ij}$ and $D_i$ is the covariant derivative compatible with the spatial metric, we can explicitly perform the Legendre transformation. 

The ``gauge condition'' (\ref{gauge}) is compatible with both cosmological and spherically symmetric backgrounds and supposing the conditions \ref{con1}-\ref{con5} listed in Introduction, we uniquely obtain the gravitational action \cite{Aoki:2020lig}
\begin{eqnarray}\label{LD-EGB}
S&=& \int dt d^3x N\sqrt{\gamma}\mathcal{L}^{\rm 4D}_{\rm EGB}
\,, \\
\mathcal{L}^{\rm 4D}_{\rm EGB} &=& \frac{M_{\rm Pl}^2}{2}
\Big[ 2R - \mathcal{M} + \frac{\tilde{\alpha} }{2} 
\Big( 8R^2 -4 R\mathcal{M} -\mathcal{M}^2
- \frac{8}{3} \big(8R_{ij}R^{ij}-4R_{ij}\mathcal{M}^{ij}
-\mathcal{M}_{ij}\mathcal{M}^{ij}\big) \Big) \Big]\,, \nonumber
\end{eqnarray} 
where $M_{\rm Pl}^2=(8\pi {G})^{-1}$ is the reduced Planck mass with $G$ being the Newton gravitational coupling constant, $R$ and $R_{ij}$ are respectively the Ricci scalar and the Ricci tensor of the spatial metric, and
\begin{eqnarray}\label{M}
\mathcal{M}_{ij} \equiv R_{ij}+\mathcal{K}^k{}_{k} \mathcal{K}_{ij}-\mathcal{K}_{ik}\mathcal{K}^k{}_{j}, 
\hspace{1cm} \mathcal{M} \equiv \mathcal{M}^i{}_{i} \,, 
\end{eqnarray}
with 
\begin{eqnarray}\label{K}
\mathcal{K}_{ij} \equiv \frac{1}{2N}( \dot{\gamma}_{ij}-2D_{(i}N_{j)}-\gamma_{ij}D^2 \lambda_{\rm GF} ) \,.
\end{eqnarray}
Here, a dot denotes derivative with respect to the time $t$ and all the effects of the constraint stemming from the ``gauge-fixing'' are now encoded in $\lambda_{\rm GF}$. The theory defined by \eqref{LD-EGB} has the time reparametrization symmetry
\begin{align} \label{time_repara}
t\rightarrow t=t(t')
\,,
\end{align}
in addition to the spatial diffeomorphism invariance \eqref{spatial-diff}. 

After formulating the theory in a consistent way, the next question is whether it is phenomenologically interesting or not and cosmology is one of the most promising setup to answer this question. Therefore, in this paper, we study cosmological implications of the theory coupled to a matter field in the form of a perfect fluid. For the sake of simplicity, we work with a minimally coupled k-essence field and the action takes the form
\begin{eqnarray}\label{action}
S = \int d^3x dt N \sqrt{\gamma} \big[\, \mathcal{L}^{\rm 4D}_{\rm EGB} + P(X) \, \big], 
\quad X \equiv g^{\mu\nu}\partial_{\mu}\phi \partial_{\nu} \phi= - \frac{1}{N^2} ( \dot{\phi} - N^i\partial_i\phi )^2 + \gamma^{ij} \partial_i\phi \partial_j\phi \,,
\end{eqnarray}
where $X$ is the kinetic term of the k-essence scalar field $\phi$ and we decomposed it by means of the $4$-dimensional ADM metric \eqref{metric}. 
The energy density, the pressure, and the sound speed of the k-essence are defined by
\begin{align}
\rho \equiv 2 X P_{, X} - P \,, \hspace{1cm}
p \equiv P \,, \hspace{1cm}
c_s^2\equiv \frac{P_{,X}}{P_{,X}+2XP_{,XX}} \,,
\end{align}
where $P_{,X}$ and $P_{,XX}$ are the first and the second derivatives of $P$ with respect to $X$, respectively. 

\subsection{Hamiltonian formalism and general discussions}

Here we summarize general discussions on our $D\to4$ Einstein-Gauss-Bonnet gravity and explain the relation to the $D=(d+1)$-dimensional Einstein-Gauss-Bonnet gravity. We also mention subtleties of the $D\to4$ limit and explain how these subtleties are resolved in our setup. (For those readers who are interested in cosmology and gravitational waves only, this subsection can be safely skipped.)

Upon performing the Legendre transformation, from (\ref{LD-EGB}) one obtains the Hamiltonian
\begin{align}
H=\int d^3x \left[ N\mathcal{H}_0(\gamma_{ij},\pi^{ij}) + N^i \mathcal{H}_i(\gamma_{ij},\pi^{ij}) + \lambda_{\rm GF} {}^3\! \mathcal{G}  \right]
, \label{4d_Hamiltonian}
\end{align}
where ${}^3\! \mathcal{G} = \sqrt{\gamma}D_k D^k \left(\frac{\pi^{ij}\gamma_{ij}}{\sqrt{\gamma}} \right)$
and the dynamical phase space variables are only $(\gamma_{ij},\pi^{ij})$ and other variables $N$, $N^i$ and $\lambda_{\rm GF}$ are regarded as the Lagrange multipliers imposing the primary constraints so that the on-shell Hamiltonian vanishes. The momentum constraint $\mathcal{H}_i$ takes the standard form and the explicit form of the Hamiltonian constraint $\mathcal{H}_0$ is given in~\cite{Aoki:2020lig}. Similarly to the shift vector $N^i$, the variable $\lambda_{\rm GF}$ appears in the action \eqref{LD-EGB} only through $\mathcal{K}_{ij}$, which ensures that the Hamiltonian is linear in $\lambda_{\rm GF}$. The theory \eqref{LD-EGB} has only two dynamical dofs since we have three first class constraints
\begin{align}
\mathcal{H}_i \approx 0
\,,
\end{align}
and a couple of second class constraints
\begin{align}
\mathcal{H}_0 \approx 0 \,, \quad {}^3\! \mathcal{G}\approx 0
\,,
\end{align}
satisfying $\{\mathcal{H}_0,  {}^3\! \mathcal{G} \} \not\approx 0$. Thus the constraint ${}^3\! \mathcal{G} \approx 0$ is second-class, meaning that the Lagrange multiplier $\lambda_{\rm GF}$ is fixed in terms of the canonical variables by the consistency condition. In principle, one can eliminate $\lambda_{\rm GF}$ from the Lagrangian or the Hamiltonian to find a description of the same theory in terms of the components of the metric $g_{\mu\nu}$ only. However, such a description is not necessarily useful for practical purposes. In particular, the equation of $\lambda_{\rm GF}$ may not be easily solved in general. It would be better to retain $\lambda_{\rm GF}$ as an independent auxiliary variable.

We should emphasise that the 4-dimensional Hamiltonian/Lagrangian does not require any assumptions about 4-dimensional spacetimes. One can use (\ref{LD-EGB}) or \eqref{4d_Hamiltonian} to analyse any 4-dimensional spacetime. Moreover, the conditions \ref{con1}-\ref{con4} listed in Introduction does not require any notion of the ``parent'' $D$-dimensional theory with $D>4$. The theory is completely defined in terms of the 4-dimensional quantities.

We also note that the Hamiltonian (\ref{4d_Hamiltonian}) is obtained through the Legendre transformation of the Lagrangian (\ref{LD-EGB}), which is based on the particular ``gauge'' (\ref{gauge}) in terms of phase space variables. In principle, at the level of the Hamiltonian the ``gauge condition'' ${}^3\! \mathcal{G} $ can be replaced by any constraint that respects the desired conditions mentioned above \cite{Aoki:2020lig}. After that, the corresponding Lagrangian can be obtained by a Legendre transformation. Indeed, in \cite{Aoki:2020lig}, the theory was originally defined in the Hamiltonian formalism and then transformed to the Lagrangian.

To understand the condition \ref{con5} and the relation to the $D$-dimensional theory, let us then briefly review how the $(3+1)$-dimensional Hamiltonian \eqref{4d_Hamiltonian} is derived from the $(d+1)$-dimensional Einstein-Gauss-Bonnet gravity (see~\cite{Aoki:2020lig} for more details). The Hamiltonian of $(d+1)$-dimensional Einstein-Gauss-Bonnet gravity is given by
\begin{align}
{}^d\! H=\int d^d x \left[ N {}^d \! \mathcal{H}_0+N^i \mathcal{H}_i \right]={}^d\! H_{\rm reg} +{}^d\! H_{\rm Weyl}
\label{D_dim}
\end{align}
where ${}^d\! H_{\rm reg}$ is the regular part of ${}^d \! H$ under the $d\to 3$ limit and ${}^d\! H_{\rm Weyl}$ is the part including the spatial Weyl tensor and the Weyl piece of $\pi^{i[k}\pi^{l]j}$ with the $1/(d-3)$ factors. 

The $d\to 3$ limit of ${}^d \! H_{\rm Weyl}$ is subtle and this is precisely the origin of subtleties of the $D\to 4$ limit\footnote{The paper \cite{Gurses:2020ofy} pointed out that $D$-dimensional Weyl tensor $W_{\mu\nu\rho\sigma}$ or more precisely the combination $W_{\mu\alpha\beta\gamma}W_{\nu}{}^{\alpha\beta\gamma}-\frac{1}{4}g_{\mu\nu} W_{\alpha\beta\gamma\delta}W^{\alpha\beta\gamma\delta}$ is problematic under the $D\to 4$ limit when the $D$-dimensional covariance is kept. On the other hand, by the use of the ADM-decomposition, it turned out that the spatial Weyl pieces are more essential for the subtleties of the limit~\cite{Aoki:2020lig}. }. As discussed in~\cite{Aoki:2020lig}, the $d\to3$ limit of ${}^d \! H_{\rm Weyl}$ is not determined by the $(3+1)$-dimensional variables only and additional fields originated from the $(d-3)$-dimensional parts of $\gamma_{ij}$ and $\pi^{ij}$, survive after taking the $d\to3$ limit of ${}^d \! H_{\rm Weyl}$. Immediately we find that \emph{the $d\to3$ limit is not unique when ${}^d \! H_{\rm Weyl}$ is retained in the Hamiltonian,}  and the resultant $D\to4$ theory depends on the specific ansatz (or symmetry) of the $D$-dimensional spacetime (for instance, the $D\to4$ theory depends on the curvature of the $(D-4)$-dimensional metric~\cite{Lu:2020iav,Kobayashi:2020wqy}). In other word, in this case, one cannot define the theory in $(3+1)$ dimensions without any attribution to the $(D-4)$-dimensional parts. Therefore, contrary to the claim of \cite{Glavan:2019inb}, \emph{the $D\to4$ limit of the Einstein-Gauss-Bonnet gravity generally involves an additional dof} (see \cite{Lu:2020iav,Kobayashi:2020wqy,Bonifacio:2020vbk})\footnote{Recall that the number of dof in the $D$-dimensional Einstein-Gauss-Bonnet gravity is $D(D-3)/2$ which is larger than 2 when $D>4$. If the theory is firstly defined in $D>4$ dimensions, there are extra dofs. One should be careful about these extra dofs while taking the $D\to 4$ limit. The results of \cite{Lu:2020iav,Kobayashi:2020wqy,Bonifacio:2020vbk} indicate that to reproduce the results of \cite{Glavan:2019inb}, at least one extra dof survives (i.e. is not decoupled) under the $D\to4$ limit.}. Apart from the non-uniqueness issue, the appearance of additional field(s) itself is not a problem of the theory. However, a crucial problem of this approach was pointed out by the papers~\cite{Kobayashi:2020wqy,Bonifacio:2020vbk}: the additional dof is infinitely strongly coupled around the FLRW background (and around the flat background) when one tries to reproduce the results of \cite{Glavan:2019inb}. This result could explain the reason why one only finds 2 dofs in the linearised equations of motion~\cite{Glavan:2019inb}. \emph{The additional dof is just infinitely strongly coupled and does not show up at the linear perturbation level.} As mentioned in Introduction, the infinite strong coupling means that the theory cannot make any physical predictions in a reliable way. Hence, this is indeed a physical problem of the naive $D\to 4$ limit of the Einstein-Gauss-Bonnet gravity. Although the issue of strong coupling is explicitly shown in the particular cases, it can be understood in a general way as follows. In \cite{Aoki:2020lig} we have shown that when we keep the Weyl part ${}^d \! H_{\rm Weyl}$ some extra dofs (at least one in the simplest case) survive after taking the limit $D\to4$. On the other hand, the apparent number of dofs at the linear level in \cite{Glavan:2019inb} is 2 as they also naively claimed that the number of dofs is 2. Then, to reproduce the result of \cite{Glavan:2019inb}, one needs to push all extra dofs to vanish which means that the setup becomes infinitely strongly coupled. In this respect, the strong coupling problem would generally arise after taking the limit if we keep Weyl part ${}^d \! H_{\rm Weyl}$. The only way to escape from this issue is to remove the Weyl term ${}^d \! H_{\rm Weyl}$ by adding appropriate counter term to the Hamiltonian. This is exactly what we investigated in \cite{Aoki:2020lig} leading to the Lorentz violating theory with which we deal in this paper. However, it is not possible to reproduce all results of \cite{Glavan:2019inb} within the setup of \cite{Aoki:2020lig}. In particular, as we will show, our setup \cite{Aoki:2020lig} does not validate the equation of motion for the tensor modes that was suggested in \cite{Glavan:2019inb}. The only way to reproduce all the results of \cite{Glavan:2019inb} is to keep the Weyl part ${}^d \! H_{\rm Weyl}$ and, as we explained above, the resultant theory inevitably suffers from the strong coupling problem. Based on these facts, the equations of motion that were suggested in \cite{Glavan:2019inb} cannot be obtained from any consistent action and in this sense the setup of \cite{Glavan:2019inb} is inconsistent.

Our approach~\cite{Aoki:2020lig} is different from above. We simply subtract the problematic term ${}^d \! H_{\rm Weyl}$ from the Hamiltonian (namely, the counter term is $-{}^d \! H_{\rm Weyl}$) in order that the $D\to4$ theory is described by the $(3+1)$-dimensional variables only. Then, by definition, there is no strong coupling problem. A price to pay here is violation of the temporal diffeomorphism invariance. Since after subtracting ${}^d \! H_{\rm Weyl}$ the Hamiltonian constraint is no longer first-class, a naive subtraction of ${}^d \! H_{\rm Weyl}$ is inconsistent~\cite{Aoki:2020lig}. We need to add an additional constraint to preserve the number of dof which we shall explain below.

In the Hamiltonian \eqref{D_dim}, the Hamiltonian constraint ${}^d \! \mathcal{H}_0 \approx 0$ is first-class. Before subtracting ${}^d \! H_{\rm Weyl}$, we then introduce the gauge-fixing condition
\begin{align}
{}^d\! H'={}^d\! H_{\rm reg} +{}^d\! H_{\rm Weyl} +\int d^d x \lambda_{\rm GF} {}^d \! \mathcal{G} 
\label{D_dim2}
\end{align}
to split the first class constraint ${}^d \! \mathcal{H}_0 \approx 0$ into a couple of second class constraints ${}^d \! \mathcal{H}_0 \approx 0,{}^d \! \mathcal{G}_0 \approx 0$. Up to here the choice of ${}^d \! \mathcal{G}$ does not lead to any physical difference at least locally. However, the theory has no explicit temporal diffeomorphism invariance due to the gauge-fixing. We then add the counter term ${}^d \! H_{\rm ct}\equiv -{}^d \! H_{\rm Weyl}$ to find a well-defined $d\to 3$ limit,
\begin{align}
{}^d\! H''={}^d\! H'+{}^d \! H_{\rm ct}= {}^d\! H_{\rm reg} +\int d^d x \lambda_{\rm GF} {}^d \! \mathcal{G} 
\,.
\label{D_dim3}
\end{align}
The originally unphysical constraint $ {}^d \! \mathcal{G}\approx 0 $ then turns to be a {\it physical} constraint since the counter term ${}^d \! H_{\rm ct}=-{}^d \! H_{\rm Weyl}$ breaks the temporal diffeomorphism invariance. The $d\to 3$ limit of \eqref{D_dim3} finally yields \eqref{4d_Hamiltonian} where the constraint ${}^3 \! \mathcal{G}\approx 0 $ of the Hamiltonian \eqref{4d_Hamiltonian} is given by the $d\to 3$ limit of the originally gauge-fixing condition $ {}^d \! \mathcal{G} \approx 0$ in $(d+1)$-dimensions. We thus call ${}^3\! \mathcal{G} \approx 0$ the ``gauge-fixing'' constraint because of its origin. Note that the constraint ${}^3\! \mathcal{G} \approx 0$ is a physical constraint to eliminate an unwanted dof in the $(3+1)$-dimensional picture. A different choice of ${}^3 \! \mathcal{G}$ gives a different $(3+1)$-dimensional theory because of the regularisation scheme that respects the spatial diffeomorphism invariance but that breaks the temporal diffeomorphism invariance.

By construction, solutions of \eqref{LD-EGB} or \eqref{4d_Hamiltonian} of the theory coincide with the $D \to 4 $ limit of solutions of the $D$-dimensional Einstein-Gauss-Bonnet gravity under the gauge condition  ${}^d\! \mathcal{G} \approx 0$ if the $D$-dimensional solutions satisfy certain conditions, namely vanishing counter term ${}^d \! H_{\rm ct}=0$. In the Hamiltonian formalism, the counter term vanishes when the spatial metric is conformally flat and the Weyl part of $\pi^{i[k}\pi^{l]j}$ vanishes. In the Lagrange formalism with the choice ${}^d\! \mathcal{G}=\sqrt{\gamma}D_k D^k(\pi^{ij}\gamma_{ij}/\sqrt{\gamma}) \approx 0$, the latter condition becomes vanishing the Weyl part of $K_{i[k}K_{l]j}$. We thus achieve the property \ref{con5} mentioned in Introduction. On the other hand, in general, some of 4-dimensional solutions of \eqref{LD-EGB} or \eqref{4d_Hamiltonian} cannot be obtained from the $D\to 4$ limit of solutions of $D$-dimensional Einstein-Gauss-Bonnet gravity. We will explicitly confirm this point by considering the cosmological solution of \eqref{action}.

Under the GR limit ${\tilde \alpha} \rightarrow 0$, $\pi^{ij}\gamma_{ij}/\sqrt{\gamma}$ is reduced to the trace of the extrinsic curvature $K\equiv K^i{}_i$. Hence, the $\tilde{\alpha} \rightarrow 0$ limit of \eqref{LD-EGB} is GR in the uniform mean curvature slice $K=K(t)$. In the present paper, we use this particular choice of the ``gauge-fixing'' constraint for the explicit calculations; however, as we will discuss, the conclusion of the present paper must be independent of ${}^3\! \mathcal{G}$.

Finally, we assume that at the classical level, the matter action respects the local Lorentz invariance. At the quantum level, the Lorentz violation in the gravity sector percolates to the matter sector via graviton loops. Such a Lorentz violation in the matter sector is suppressed not only by $\tilde{\alpha}$ but also by a negative power of $M_{\rm Pl}^2$ and thus is under control. Phenomenological implications of the Lorentz violation induced by graviton loops in the matter sector deserves further investigation but they are beyond the scope of the present paper.

\section{Cosmological background}
\label{sec_background}

In this section we present the cosmological background equations in spatially flat FLRW spacetime driven by a homogeneous k-essence field 
\begin{equation}\label{metric-BG}
N = \bar{N}(t) \,, \hspace{1cm} N^i = 0 \,, \hspace{1cm} \gamma_{ij} = a(t)^2 \delta_{ij} \,, \hspace{1cm}
 \phi = \bar{\phi}(t)\,,
\end{equation}
where $a(t)$ is the scale factor. We shall put a bar to represent the background quantities, e.g.~$\bar{\phi}$ is the background value of $\phi$.  Note that the FLRW ansatz is compatible with the time reparametrization \eqref{time_repara}. Considering the spatially homogeneous ansatz $\lambda_{\rm GF} = \bar{\lambda}_{{\rm GF}}(t)$, the ``gauge-fixing'' term in \eqref{K} vanishes and ${\cal K}_{ij}$ reduces to the standard extrinsic curvature for the FLRW background
\begin{equation}\label{extrinsic-curvature}
{\cal K}^i{}_j = H \delta^i{}_j \,,
\hspace{1cm} H \equiv \frac{{\dot a}}{{\bar N}a} \,.
\end{equation}
The background value of $\lambda_{\rm GF}$ is undetermined by equations of motion because the ``gauge condition'' that we have chosen does not fix the freedom of the time reparametrization symmetry \eqref{time_repara}.

Substituting \eqref{metric-BG} and \eqref{extrinsic-curvature} into the action \eqref{action}, we find the following homogeneous and isotropic minisuperspace action
\begin{equation}\label{action-BG}
{\bar S} = V_0 \int dt {\bar N} a^3 \left[
P\big( {\bar X} \big) - 3 M_{\rm Pl}^2 H^2 
- {\tilde \alpha} M_{\rm Pl}^2 H^4 \right] \,, 
\hspace{1cm}
{\bar X} = - \Big( \frac{\dot{\bar{\phi}}}{\bar N}\Big)^2 \,,
\end{equation}
where $V_0=\int d^3x$ is the spatial volume and we assume it to be large enough but finite.

Hereinafter,  after taking the variations to obtain the background equations, we set ${\bar N}=1$ by the use of the time reparametrization symmetry \eqref{time_repara}. The dot simply means the time derivative with respect to the cosmic time.
Varying the above action with respect to the lapse function, we find the first Friedmann equation
\begin{eqnarray}\label{Friedmann-1}
3 M_{\rm Pl}^2\left( H^2 +\tilde{\alpha} H^4 \right) = \bar{\rho} \,. 
\end{eqnarray}
Taking variation with respect to the scale factor gives the second Friedmann equation, which after using \eqref{Friedmann-1} simplifies to
\begin{align}\label{Friedmann-2}
-2 M_{\rm Pl}^2 \Gamma \dot{H} &= \bar{\rho}+\bar{p} \,.
\end{align}
Here, we have defined the function 
\begin{align}
  \Gamma &\equiv 1+2{\tilde \alpha}H^2 \,, \label{def_Gamma}
\end{align}
as in \cite{Glavan:2019inb} to make the comparison of the results easy. 
Finally, variation of \eqref{action-BG} with respect to the k-essence field gives the equation of local conservation of the stress-energy tensor,
\begin{equation}\label{KG}
\dot{\bar{\rho}} + 3 H (\bar{\rho} + \bar{p}) = 0\,.
\end{equation}
One can of course derive the same set of background equations of motion by simply expand the action \eqref{action} up to linear order in general perturbations. The equation \eqref{Friedmann-2} can be derived from \eqref{Friedmann-1} and \eqref{KG} due to the time reparametrization symmetry \eqref{time_repara}.

Although we considered the shift symmetric k-essence scalar as a matter field for the sake of simplicity, the results \eqref{Friedmann-1}, \eqref{Friedmann-2} and \eqref{KG} are also applicable for a more general perfect fluid. These results coincide with those obtained in \cite{Glavan:2019inb}. However, the coincidence will not always happen for other types of solutions. It happened here since the spatial part of the FLRW metric \eqref{metric-BG} is conformally flat, thanks to the property \ref{con5} of the consistent $4$-dimensional theory considered in the present paper. More precisely, in \cite{Aoki:2020lig}, we have shown that the ambiguities of the limit $D\to4$ are originated from the counter terms that cancel the divergences due to the Weyl parts of the spatial Riemann tensor $R_{ijkl}$ and tensor ${\cal M}_{ijkl} = R_{ijkl}+2{\cal K}_{i[k} {\cal K}_{l]j}$. In FLRW spacetime, the Weyl pieces of both of these tensors vanish which is clear from \eqref{metric-BG} and \eqref{extrinsic-curvature}. Therefore the subtleties will not arise in this special case. That is the reason why the equations \eqref{Friedmann-1}-\eqref{KG} coincide with those in the naive (and inconsistent) $D\rightarrow 4$ limit studied in \cite{Glavan:2019inb} (see also \cite{Narain:2020qhh} for the minisuperspace action). The same happens for the spherically symmetric spacetime. In particular, the black hole solution found in \cite{Glavan:2019inb} is also a vacuum solution of our theory with Lagrangian density \eqref{LD-EGB} as one can easily confirm.

\section{Scalar perturbations}
\label{sec_scalar}

Having studied background equations, we now consider scalar perturbations around the background geometry \eqref{metric-BG} as
\begin{equation}\label{metric-scalar}
N = 1 + A \,,\quad  N^i = \delta^{ij} \partial_j B \,, \quad 
\gamma_{ij} = a(t)^2 \big[ ( 1 + 2 \psi ) \delta_{ij} + \partial_i\partial_jE \big] \,, \quad 
\phi = \bar{\phi}(t) + \delta \phi\,. 
\end{equation}
We should also consider scalar perturbations in the ``gauge-fixing'' term as $\lambda_{\rm GF} = {\bar \lambda}_{\rm GF} + a^2 \delta\lambda$. We therefore deal with six scalar variables $(A,B,\psi,E,\delta\lambda,\delta\phi)$. The theory is invariant under the infinitesimal version of the spatial diffeomorphism \eqref{spatial-diff}, $x^i \to x^i + \xi^i$, and using the usual decomposition $\xi^i = \delta^{ij}\partial_j\xi$ in terms of a spatial scalar $\xi$, we can set $E=0$ in \eqref{metric-scalar} by fixing this gauge freedom.

Expanding the action up to the quadratic order in perturbations and using the background equations, we obtain the quadratic action in terms of $(A,B,\psi,\delta\lambda,\delta\phi)$.
We then change the perturbation variables $(\psi,A,\delta \phi)$ into the ``gauge-invariant'' combinations
\begin{eqnarray} \label{gauge_inv}
\Psi \equiv \psi + a^2 H B \,, \hspace{1cm}
\Phi \equiv A + a^2 \dot{B} + 2 a^2 H B \,, \hspace{1cm}
\delta\phi_{\rm inv} \equiv \delta\phi + a^2 \dot{\bar{\phi} } B \,,
\end{eqnarray}
which are invariant under the linearised temporal diffeomorphism if the theory has such a symmetry. Although the action \eqref{LD-EGB} has no temporal diffeomorphism invariance, the change of the variables, $(\psi,A,\delta \phi) \to (\Psi,\Phi, \delta\phi_{\rm inv} )$, is quite useful. After changing the variables in this way, the shift perturbation $B$ plays the role of a Lagrange multiplier to implement the constraint
\begin{align}
\delta \lambda=0
\,.
\end{align}
Substituting this solution to the quadratic action, we obtain the action in terms of $(\Psi, \Phi, \delta\phi_{\rm inv})$.

The variable $\Phi$ is also non-dynamical and can be integrated out. We further define the new variable
\begin{equation}\label{zeta}
\zeta \equiv \psi-\frac{H}{\dot{\bar{\phi}}} \delta \phi= \Psi-\frac{H}{\dot{\bar{\phi}}} \delta\phi_{\rm inv}
\,,
\end{equation}
and then integrate out $\Psi$. As a result, the quadratic action in Fourier space takes the form\footnote{We represent the amplitude of the Fourier transformations with the same notation of the corresponding field in the real space.}
\begin{eqnarray}\label{S2-SS-Fourier}
S^{\rm SS}_{2} = M_{\rm Pl}^2 \int dt d^3k\, a^3 \frac{ \epsilon \Gamma}{c_s^2}
\left[ \dot{\zeta}^2 -\frac{c_s^2 k^2}{a^2} \zeta^2 \right]
\,,
\end{eqnarray}
where $\epsilon=-\dot{H}/H^2$ and $\Gamma$ is defined by \eqref{def_Gamma}. The equation of motion of $\zeta$ is
\begin{align}
\ddot{\zeta}+3H\left( 1+ \frac{ (\epsilon/c_s^2)^{\cdot}}{3H \epsilon/c_s^2} -\frac{4\tilde{\alpha} \epsilon H^2}{3 \Gamma} \right)\dot{\zeta} + \frac{c_s^2 k^2}{a^2}\zeta=0
\,,
\end{align}
which agrees with that in \cite{Glavan:2019inb} when $c_s=1$. The coincidence with \cite{Glavan:2019inb} is due to the fact that the spatial metric of scalar perturbations \eqref{metric-scalar} is conformally flat and that the Weyl part of $K_{i[k}K_{l]j}$ vanishes at the linear order of the perturbations. As long as the null energy condition for the perfect fluid $\rho+p\geq0$ holds, $\dot{H}\leq 0$ and the scalar mode is free of any pathologies. We therefore can implement it to construct an inflationary scenario by considering standard slow roll potential. We can also construct a late time scalar-tensor scenario where the k-essence may play the role of dark energy.

It would be worth emphasizing that we have $\delta \lambda =0$ from the equations of motion and thus $\delta\lambda$ does not affect the analysis of scalar perturbations at all after we integrate out $\delta\lambda$. This implies that we can obtain exactly the same result of the scalar perturbations even if the original action \eqref{LD-EGB} has no ``gauge-fixing'' term $\lambda_{\rm GF}$. Indeed, studying the scalar perturbations based on \eqref{LD-EGB} \emph{without} $\lambda_{\rm GF}$ by the use of the variables \eqref{gauge_inv}, one can find that the quadratic action is independent of $B$, meaning the invariance under the linearised temporal diffeomorphism. The emergence of the linearised symmetry may be understood by the fact that the spatial metric \eqref{metric-scalar} is conformally flat and that the Weyl part of $K_{i[k}K_{l]j}$ vanishes. The quadratic order action of \eqref{LD-EGB} without $\lambda_{\rm GF}$ under the ansatz \eqref{metric-scalar} may be obtained by taking a naive $D\rightarrow 4$ limit of that of the $D$-dimensional Einstein-Gauss-Bonnet gravity for which there exists the temporal diffeomorphism invariance. However, one should recall that the full $4$-dimensional theory without $\lambda_{\rm GF}$ has neither the temporal diffeomorphism invariance nor an additional constraint that would remove an unwanted degree in the phase space and thus is inconsistent as shown in \cite{Aoki:2020lig}, contrary to the consistent theory \eqref{LD-EGB} with $\lambda_{\rm GF}$. One has to work based on \eqref{LD-EGB} \emph{with} $\lambda_{\rm GF}$ and obtain $\delta \lambda=0$ from the equations of motion at the level of linear perturbations. Nonetheless, this observation implies that the result of the scalar perturbations is not affected by the specific form of the ``gauge-fixing'' constraint when we use the gauge-invariant variables \eqref{gauge_inv}. The results here must be robust against changes of the choice of the ``gauge-fixing''. The independence of the ``gauge-fixing'' constraint must be true even at the cubic order action because we can still use the linear order solution $\delta \lambda=0$ to compute the cubic action. On the other hand, at the quartic order, a different choice of the ``gauge-fixing'' constraint would give a different result.

\section{Vector perturbations}
\label{sec_vector}

We then briefly discuss the vector type perturbations. The theory \eqref{LD-EGB} has only tensorial dofs. Since the perfect fluid described by the k-essence field has only scalar dof, there is not any vectorial mode in the matter sector to source the gravitational vectorial dofs. Consequently, there must be no dynamical dofs in the vector sector of the theory. By explicit calculations, we have also confirmed that the vector type perturbations are not dynamical and, therefore, we do not discuss them further.

\section{Tensor perturbations}
\label{sec_tensor}

Of more interests in the consistent $D\to4$ Einstein-Gauss-Bonnet gravity is the tensor perturbations. The tensor perturbations around the background geometry \eqref{metric-BG} are given by
\begin{equation}\label{metric-tensor}
N = 1\,, \quad N^i = 0\,, \quad \gamma_{ij} = a(t)^2 ( \delta_{ij} + h_{ij} ) \,, \quad
 \phi = \bar{\phi}(t)\,,
\end{equation}
where $h_{ij}$ represents tensor perturbations satisfying the transverse traceless condition $\partial^i h_{ij} = 0 = h^{i}{}_{i}$ and we have considered $\bar{N}=1$ so that $t$ is the cosmic time.

Substituting \eqref{metric-tensor} into the action \eqref{action} and performing some integrations by part, we find the quadratic action for the tensor perturbations as follows
\begin{eqnarray}\label{S2-TT-real}
S^{\rm TT}_{2} = 
\frac{M_{\rm Pl}^2}{8} \int dt d^3x a^3 
\bigg[ \Gamma\bigg(\dot{h}_{ij} \dot{h}^{ij} 
- \frac{c_T^2}{a^2}\partial_k h^{ij} \partial^k h_{ij} \bigg)
- \frac{4{\tilde \alpha}}{a^4} \partial_l \partial_k h^{ij} 
\partial^l \partial ^kh_{ij} \bigg] \,,
\end{eqnarray}
or
\begin{eqnarray}\label{S2-TT-Fourier}
S^{\rm TT}_{2} = \frac{M_{\rm Pl}^2}{8} \int dt d^3k a^3 \Gamma 
\bigg[ \dot{h}_{ij} \dot{h}^{ij} - \bigg( \frac{c_T^2k^2}{a^2} 
+ \frac{4{\tilde \alpha}}{\Gamma} \frac{k^4}{a^4} \bigg) h_{ij} h^{ij}\bigg] \,,
\end{eqnarray}
in the Fourier space, where 
\begin{equation}\label{cs2-T}
c_T^2 \equiv 1+ \frac{ \dot{\Gamma}}{H\Gamma} = 1-\frac{4\tilde{\alpha}}{\Gamma} \epsilon H^2\,.
\end{equation}
Its equation of motion is
\begin{align}
\ddot{h}_{ij}+3H \left( 1- \frac{4\tilde{\alpha} \epsilon H^2}{3 \Gamma} \right) \dot{h}_{ij} 
+ \left( \frac{c_T^2 k^2}{a^2} + \frac{4{\tilde \alpha}}{\Gamma} \frac{k^4}{a^4} \right) h_{ij}=0
\,.
\end{align}
Although $c_T^2$ can be negative, there exists the $k^4$ term coefficient of which is positive so long as $\tilde{\alpha}>0$. Hence, for ${\tilde \alpha}>0$ and $\dot{H}<0$ the tensor modes are free of either ghost or gradient instabilities. If $c_T^2<0$ but $\tilde{\alpha}>0$, only the IR modes of the tensor perturbations are unstable which is not necessarily pathological, similarly to the Jeans instability \cite{ArkaniHamed:2003uy,ArkaniHamed:2005gu,Creminelli:2006xe,Gumrukcuoglu:2016jbh}.

Interestingly, tensor modes have a modified dispersion relation in this scenario like the Ho\v{r}ava-Lifshitz gravity. Except the $k^4$ term, the equation of motion of the tensor mode coincides with that derived from the naive (and inconsistent) $D\to 4$ limit in \cite{Glavan:2019inb}. This result reveals that the result in \cite{Glavan:2019inb} only captures the IR limit of the consistent theory \eqref{LD-EGB}, but the consistency of the $4$-dimensional description of the $D\rightarrow 4$ theory without the strong coupling requires the $k^4$ term in the dispersion relation of the gravitational waves which significantly changes the physics in the UV regime.

The appearance of the $k^4$ term can be understood if one recalls that, in order to have a consistent theory of $D\to4$ Einstein-Gauss-Bonnet gravity with two dofs, one needs to introduce counter terms to cancel divergences due to the spatial Weyl pieces of the Gauss-Bonnet term as a regularisation of the Hamiltonian or/and the action \cite{Aoki:2020lig}. These counter terms change the second-order structure of the Gauss-Bonnet term so that higher spatial derivative term are allowed (and indeed present) in this theory. Since we defined the theory in such a way that the number of dofs is two at fully nonlinear orders, higher time derivatives are forbidden by construction. However, the spatial higher derivatives are not forbidden and that is the reason why we found a spatial higher derivative term in \eqref{S2-TT-real} or equivalently a $k^4$ term in \eqref{S2-TT-Fourier}. As we have seen, the higher order spatial derivative does not appear in the scalar perturbation sector since the Weyl pieces are traceless and can only modify the second order structure of the tensorial modes. Appearance of the $k^4$ correction reveals that the small scale structure of the tensor sector of the theory \eqref{action} is different from the standard GR and even from many modified gravity theories. According to the classification of \cite{Aoki:2018brq}, the theory \eqref{LD-EGB} is a type-II minimally modified gravity, i.e. a theory without the Einstein frame, since the dispersion relation is no longer the same as GR.

From \eqref{cs2-T} we also see that the IR speed of gravitational waves is corrected by the Gauss-Bonnet term. The current bound on the speed of gravitational waves is $|1-c_T|\lesssim10^{-15}$ \cite{Monitor:2017mdv}, which would lead to a rather weak upper bound $\tilde{\alpha} \lesssim 10^{50}\,{\rm eV}^{-2}$ on the rescaled Gauss-Bonnet coupling constant if we ignore the effects of $k^4$ term. A similar bound can be obtained from the correction that appears in the friction term \cite{Saltas:2014dha,Belgacem:2019pkk}~\footnote{We thank Charles Dalang for bringing this point to our attention.} if we again ignore the effects of $k^4$ term. Actually, the $k^4$ term cannot be ignored if $\tilde{\alpha}$ is as large as $\sim 10^{50}\,{\rm eV}^{-2}$. Taking into account the $k^4$ term, therefore, much stronger bound then arise from the gravitational waves observational bound \cite{Abbott:2017vtc,Sotiriou:2017obf,Gumrukcuoglu:2017ijh} as
\begin{equation}\label{bound}
\tilde{\alpha} \lesssim (10\, \mbox{meV})^{-2} \,.
\end{equation}

Moreover, in the IR limit, the additional terms in the equations of motion due to the $D\to 4$ Gauss-Bonnet term should be smaller than those due to the Einstein-Hilbert term when we apply the theory to physical systems that have already been observationally/experimentally confirmed to reproduce GR predictions. Following the method investigated in \cite{Allahyari:2020jkn}, this criterion implies ${\tilde \alpha}{\cal R}\leq1$, where ${\cal R}$ is an average Gaussian curvature which can be identified with nonzero tetrad components of the Riemann tensor for the system under consideration. Around a compact astronomical object, the background geometry is approximately given by the Schwarzschild-type solution. On the surface of a compact astronomical object we then have ${\cal R} \sim r_{\rm S}/r^3$, where $r_{\rm S} = M/(4\pi{M}_{\rm Pl}^2)$ is the Schwarzschild radius of the object with mass $M$ and radius $r$ \cite{Allahyari:2020jkn}. In the case of a neutron star, the typical values of the mass and radius are $M = 10^{66}$ eV and $r = 5\times 10^{10}$ eV$^{-1}$ which gives the bound ${\tilde \alpha} \lesssim 10^{22}\,  {\rm eV}^{-2}$. This is much weaker than what we found from the $k^4$ term in the dispersion relation of the gravitational waves. Hence, the conservative bound on $\tilde{\alpha}$ is typically of the order of meV$^{-2}$ which reads $\tilde{\alpha} M_{\rm pl}^2 \lesssim 10^{58}$ in terms of the dimensionless combination.

\section{Summary and discussions}\label{summary}

Very recently, we proposed a consistent $4$-dimensional theory of gravity with two dofs that serves as a consistent realization of $D\to 4$ Einstein-Gauss-Bonnet gravity \cite{Aoki:2020lig}. The consistent theory is different than the previously suggested, naive (and inconsistent) $D\to 4$ limit so that it can validate some of the claims, but not all, of \cite{Glavan:2019inb} in a consistent manner. Contrary to the scalar-tensor descriptions of $D\to 4$ Einstein-Gauss-Bonnet gravity~\cite{Lu:2020iav,Kobayashi:2020wqy,Bonifacio:2020vbk,Fernandes:2020nbq,Hennigar:2020lsl}, our theory has no strong coupling problem and provides a novel gravitational theory with only two dynamical dofs while the temporal diffeomorphism invariance is broken. As argued in \cite{Aoki:2020lig}, the action of the consistent theory \eqref{LD-EGB} is uniquely determined by requiring the conditions \ref{con1}-\ref{con5} in the $4$-dimensional spacetime, up to a choice of a constraint that stems from the temporal gauge condition. Due to the violation of the temporal part of the $4$-dimensional general covariance, our theory does not contradict with the Lovelock theorem.

In the present paper, we studied cosmological implications of this theory in the presence of a minimally coupled k-essence field. All results can be straightforwardly translated into the case of a perfect fluid. We studied linear perturbations around a spatially flat FLRW background and explicitly confirmed that there only exists two gravitational dofs. We also showed that for $\tilde{\alpha}>0$ and $\dot{H}<0$, all modes are free of any pathologies so that one can construct early and/or late time cosmological models in this framework. The tensor perturbations are modified from those in GR so the IR speed of gravitational waves gets modifications from the Gauss-Bonnet term and, more interestingly, a $k^4$ term shows up in the dispersion relation. The former result is already found in \cite{Glavan:2019inb} while the latter is the specific characteristic of the consistent theory of the $D\to 4$ Einstein-Gauss-Bonnet gravity with two dofs. Taking into account the $k^4$ term, observational bounds on the propagation of gravitational waves then gives the bound $\tilde{\alpha} \lesssim (10\, \mbox{meV})^{-2}$ on the rescaled Gauss-Bonnet coupling constant\footnote{Other constraints on $\tilde{\alpha}$ can be obtained from the inflationary universe. However, additional scalar has to be introduced to cause and end the inflation in the $D\to 4$ Einstein-Gauss-Bonnet gravity so the constraints from the inflation should be model dependent. We only focus on the model-independent constraints in this paper and we leave model-dependent constraints for our future study.}.

The theory is uniquely defined up to a choice of the constraint ${}^3\!\mathcal{G} \approx 0$ that stems from a gauge-fixing condition. However, we have argued that the analysis of the linear perturbations are indeed independent of the choice of this constraint. The vector and the tensor modes would not be affected since the constraint stems from the temporal gauge fixing. For the scalar modes, at the linear level, the spatial metric is conformally flat. Therefore, the scalar part would not be affected by the choice of the additional constraint as well and we have explicitly confirmed this fact. However, the constraint stemming from the temporal gauge fixing may contribute at non-linear orders, e.g.~to the tri-spectrum of 4-point functions.

We have seen that background equations and linear scalar modes are the same as those obtained from a naive (and inconsistent) $D\to 4$ limit in \cite{Glavan:2019inb}. On the other hand, the appearance of the higher order spatial derivatives in the tensor perturbations clearly shows that our result \eqref{S2-TT-real} for tensor modes is completely different than those obtained in \cite{Glavan:2019inb} in the UV regime. The linear tensor perturbations analysis presented in \cite{Glavan:2019inb} cannot be realized in a consistent nonlinear theory, especially in the UV regime. In this sense, the present paper clarifies that some of the results in \cite{Glavan:2019inb} can be reproduced by the use of the consistent $4$-dimensional theory \eqref{LD-EGB} instead of the questionable (and ill-defined) $D\rightarrow 4$ limit, but taking the naive $D\rightarrow 4$ limit of $D$-dimensional solutions is inconsistent in general.

The situation is similar for the case of the black hole solution presented in \cite{Glavan:2019inb}. It can be easily checked that the black hole solution presented in \cite{Glavan:2019inb} is also a solution of the consistent theory with the action \eqref{LD-EGB} while it is not clear whether analysis of the quasinormal modes \cite{Konoplya:2020bxa,Churilova:2020aca,Mishra:2020gce,Zhang:2020sjh,Aragon:2020qdc,Churilova:2020mif,Devi:2020uac,Liu:2020lwc} for the naive (and inconsistent) setup of \cite{Glavan:2019inb} are still valid in the consistent theory with Lagrangian density \eqref{LD-EGB}. In particular, analysis of tensor quasinormal modes \cite{Konoplya:2020bxa} would most probably change because of the modification of the dispersion relation. It is also expected that rotating black holes in the consistent $4$-dimensional theory and the naive $D\to 4$ limit of those in $D$-dimensional Einstein-Gauss-Bonnet theory may be different since the Kerr spacetime (in the $\tilde{\alpha}\to 0$  limit) does not admit conformally flat spatial sections~\cite{Garat:2000pn,deFelice:2019hwo}.

The consistent theory \eqref{LD-EGB} may open a new window of opportunity for modified theories of gravity. The dispersion relation of the tensor modes takes the form $\omega^2=c_T^2 k^2 + \beta k^4/M_*^2$, where the $k^4$ term originates from the ``Gauss-Bonnet term''. Appearance of the $k^4$ term makes the theory different than not only general relativity but also many modified gravity theories at small scales. Although the behaviour of the dispersion relation is in a sense similar to the Ho\v{r}ava-Lifshitz gravity, the theory has no scalar graviton and has only tensor modes. It is thus interesting to look for phenomenological implications of the theory \eqref{LD-EGB} further.

Furthermore, one can easily generalize the action \eqref{LD-EGB} by removing the condition(s) \ref{con4} and/or \ref{con5}. For instance, if one drops the condition \ref{con4}, terms like the spatial Cotton tensor squared is allowed. It is then expected that the dispersion relation of gravitational waves should acquire a $k^6$ term just like the Ho\v{r}ava-Lifshitz gravity. As already commented in \cite{Aoki:2020lig}, a more general theory satisfying the conditions \ref{con1}-\ref{con3} may be given by
\begin{align} \label{general_S}
S=\int dt d^3x N \sqrt{\gamma}\mathcal{L}(\gamma_{ij}, \mathcal{K}_{ij}, R_{ij}, D_i)
\,,
\end{align}
where $\mathcal{K}_{ij}$ is defined by \eqref{K}. The structure of the Lagrangian and $\mathcal{K}_{ij}$ ensures that $N$ and $\lambda_{\rm GF}$ are Lagrange multipliers in the Hamiltonian, meaning the existence of a pair of second-class constraints, i.e. the Hamiltonian constraint $\mathcal{H}_0(\gamma_{ij},\pi^{ij})\approx 0$ and the other stemming from the gauge-fixing condition ${}^3\! \mathcal{G}=\sqrt{\gamma}D_k D^k (\pi^{ij}\gamma_{ij}/\sqrt{\gamma}) \approx 0$. In the case of the gauge-fixed GR, we have $\{\mathcal{H}_0, {}^3\! \mathcal{G}\}\not\approx 0$. Hence, the conditions \ref{con1}-\ref{con3} hold as long as $\mathcal{L}$ is a spatial scalar and has a smooth GR limit in the parameter space of the theory. One may use \eqref{general_S} as a general framework of the $4$-dimensional gravity with two gravitational dofs, namely minimally modified gravity. We leave further studies on \eqref{general_S} as well as \eqref{LD-EGB} for future works.

\vspace{0.7cm}

{\bf Acknowledgments:} 
K.A. and M.A.G. acknowledge the xTras package~\cite{Nutma:2013zea} which was used for tensorial calculations. The work of K.A. was supported in part by Grants-in-Aid from the Scientific Research Fund of the Japan Society for the Promotion of Science, No.~19J00895 and No.~20K14468. The work of M.A.G. was supported by Japan Society for the Promotion of Science Grants-in-Aid for international research fellow No. 19F19313. The work of S.M. was supported in part by Japan Society for the Promotion of Science Grants-in-Aid for Scientific Research No.~17H02890, No.~17H06359, and by World Premier International Research Center Initiative, MEXT, Japan.

{}

\end{document}